\documentclass[epj]{svjour}
\usepackage{amsfonts,amssymb,amsmath}
\usepackage{dcolumn,epsfig,color}

\begin{document}

\title{An update on fine-tunings in the triple-alpha process}

{\color{red}
\author{Timo A. L\"{a}hde \inst{1} 
\and Ulf-G.~Mei{\ss}ner\inst{2,1,3}
\and Evgeny Epelbaum\inst{4}  
%
}                     
%
%
\institute{
Institut~f\"{u}r~Kernphysik,~Institute~for~Advanced~Simulation and
J\"{u}lich~Center~for~Hadron~Physics, \\ 
Forschungszentrum~J\"{u}lich, D-52425~J\"{u}lich,~Germany
\and Helmholtz-Institut~f\"{u}r~Strahlen-~und~Kernphysik~and~Bethe~Center~for
Theoretical~Physics, Universit\"{a}t~Bonn, \\ D-53115~Bonn,~Germany
\and Tbilisi State University, 0186 Tbilisi, Georgia
\and Ruhr-Universit\"at Bochum, Fakult\"at f\"ur Physik und Astronomie,
Institut f\"ur Theoretische Physik II,\\ 
  D-44780 Bochum, Germany 
}
}
\date{Received: date / Revised version: date}
%

\abstract{
The triple-alpha process, whereby evolved stars create carbon and oxygen, is believed to be fine-tuned to a high degree. Such
fine-tuning is suggested by the unusually strong temperature dependence of the triple-alpha reaction rate at stellar temperatures.
This sensitivity is due to the resonant character of the triple-alpha process, which proceeds through the so-called ``Hoyle state'' of $^{12}$C with
spin-parity $0^+$. The question of fine-tuning can be studied within the {\it ab initio} framework of nuclear
lattice effective field theory, which makes it possible
to relate {\it ad hoc} changes in the energy of the Hoyle state to changes in the fundamental parameters of the
nuclear Hamiltonian, which are the light quark mass $m_q$ and the electromagnetic fine-structure constant. Here, we update the 
effective field theory calculation of the sensitivity of the triple-alpha process to small changes in the fundamental parameters. In particular,
we consider recent high-precision lattice QCD calculations of the nucleon axial coupling $g_A$, as well as new and more comprehensive results from
stellar simulations of the production of carbon and oxygen. While the updated stellar simulations allow for much larger {\it ad hoc} shifts
in the Hoyle state energy than previously thought, recent lattice QCD results for the nucleon S-wave singlet 
and triplet scattering lengths now disfavor the scenario of no fine-tuning in the light quark mass $m_q$.
} 

\maketitle

\section{Introduction}

The production of carbon and heavier elements in stars is complicated by the instability of the $^8$Be nucleus.
The way this bottleneck is circumvented in nature is by means of the triple-alpha process, where the production rate
of $^{12}$C is strongly enhanced by a fortuitously placed $0^+$ resonance, known as the Hoyle state~\cite{Hoyle:1954zz}.
As small ad hoc changes
in the excitation energy of the Hoyle state relative to the triple-alpha threshold can lead to large changes in the relative abundance 
of carbon and oxygen, the question arises whether the universe should be regarded as fine-tuned with respect to the likelihood of
carbon-oxygen based life to arise, for a recent review on fine-tunings  see~\cite{Adams:2019kby}.
The physics of the triple-alpha process has recently been studied using nuclear lattice effective field theory (NLEFT).
The ground state energies of $^4$He, $^8$Be and $^{12}$C, and of the energy of the Hoyle state in $^{12}$C,
were all found to be strongly correlated with respect to small changes in the fundamental constants of nature, an
effect of the clustering of alpha particles in the respective nuclei.
We review here how the sensitivity of the triple-alpha reaction rate with respect to small changes in the light quark mass
and the electromagnetic fine-structure constant is treated in the effective field theory (EFT) framework.
The main source of uncertainty
is due to the short-range part of the nucleon-nucleon interaction, and we discuss recent progress in narrowing down this uncertainty
using updated theoretical knowledge of the quark mass dependence of the two-nucleon S-wave 
scattering parameters, including the results of recent lattice QCD work. We also contrast this theoretical treatment with recent
high-precision calculations of stellar nucleosynthesis, which find that the allowable range of Hoyle state energies is larger than previously
thought~\cite{Huang:2018kok}.

This paper is structured as follows. In Sec.~\ref{sec_Ham} we  update the pion (quark) mass dependence of the nuclear
Hamiltonian, which is central for the following discussion. In Sec.~\ref{sec_stellar} we review the
current status of stellar nucleosynthesis calculations, with focus on the resulting abundances of carbon and oxygen under 
\textit{ad hoc} shifts in the Hoyle state resonance. In particular, we pay attention to recent new results in this  field.
In Sec.~\ref{sec_sensitivity}, we revisit the theoretical status
of EFT calculations of the sensitivity of the Hoyle state energy to small changes in the light quark mass $m_q^{}$ 
and the electromagnetic fine-structure constant $\alpha_\mathrm{em}$. Finally, in Sec.~\ref{sec_discussion} we discuss
how the EFT treatment of the triple-alpha process could be improved with regards to recent progress in the nuclear lattice 
EFT description of the nuclear forces.


\section{Quark mass dependence of the nuclear Hamiltonian \label{sec_Ham}}

The ground-state energies and spectra of light and medium-mass nuclei can be calculated to a good precision
in the framework of NLEFT, as described in detail in the monograph~\cite{Lahde:2019npb}. Variations of the
fundamendat parameters like the average light quark mass or the electromagnetic fine-structure constant can
also be investigated within this approach. In what follows, we update our knowledge of the quark mass
dependence of the nuclear Hamiltonian, which is central to the study of fine-tunings in the triple-alpha process. 
Note that to high accuracy the Gell-Mann-Oakes-Renner relation $M_\pi^2 \sim m_q$, with $m_q =(m_u+m_d)/2$ the
average light quark mass, is fulfilled in QCD and we thus can use the notions ``quark mass dependence'' and ``pion
mass dependence'' synonymously. This update concerns in particular the hadronic parameters $x_1$ and $x_2$ 
and the nuclear parameters
$\bar{A}_s$ and $\bar{A}_t$. For details, the reader is referred to Refs.~\cite{Epelbaum:2012iu,Epelbaum:2013wla,Lahde:2019npb}.

We first discuss $x_1$, which describes the dependence of the nucleon
mass $m_N$ on the pion mass $M_\pi$.
This is related to the pion-nucleon $\sigma$-term $\sigma_{\pi N}$, via
\begin{equation}
x_1^{} = 2\frac{\sigma_{\pi N}^{}}{M_\pi^{}},
\end{equation}
and we note that the best determinations of $\sigma_{\pi N}$ are from the recent Roy-Steiner-equation analyses of pion-nucleon 
scattering, leading to $\sigma_{\pi N} = (59.1 \pm 3.5)$~MeV~\cite{Hoferichter:2015dsa} (with the inclusion of pionic hydrogen and deuterium data)
and $\sigma_{\pi N} = (58 \pm 5)$~MeV~\cite{RuizdeElvira:2017stg} (pion-nucleon scattering data only). We take the
central value of Ref.~\cite{Hoferichter:2015dsa} and the uncertainty of Ref.~\cite{RuizdeElvira:2017stg},
to be on the conservative side.  While this gives
\begin{equation}
x_1 = 0.84(7), 
\label{eq:x1val}
\end{equation}
we note that lattice QCD determinations give systematically smaller values for $\sigma_{\pi N}$ and thus $x_1$. 
For reasons explained in
Ref.~\cite{Hoferichter:2016ocj}, such as the inconsistency of the lattice QCD values with the precisely determined
S-wave pion-nucleon scattering lengths, we do not consider the lattice QCD results  here.

Next, we turn to $x_2$, which describes the dependence of the strength
of the one-pion
exchange (OPE)  $g_A/(2F_\pi)$ on $M_\pi$. This
is given by
\begin{equation}
x_2^{} =
\frac{1}{2 F_\pi^{}} 
\left.\frac{\partial g_A^{}}{\partial M_\pi^{}}\right|_{M_\pi^\mathrm{ph}}
- \frac{g_A^{}}{2 F_\pi^2} 
\left.\frac{\partial F_\pi^{}}{\partial M_\pi^{}}\right|_{M_\pi^\mathrm{ph}},
\label{eq:x2}
\end{equation}
with $M_\pi^\mathrm{ph}$ denoting the physical value of the pion mass. Noe that $x_2^{}$ turned out to
be small and of indeterminate sign in Ref.~\cite{Epelbaum:2013wla}. 
This was largely due to the inconclusive situation of lattice QCD calculations of $g_A$. Such problems have recently
been overcome by high-precision lattice QCD calculations with close-to-physical quark masses~\cite{Chang:2018uxx}. In particular,
consistent values of $g_A$ with minimal model dependence were obtained for a range of polynomial and
chiral perturbation theory (ChPT) extrapolations in $M_\pi$. In order to make use of the analysis of Ref.~\cite{Chang:2018uxx}, we define
\begin{equation}
\frac{\partial g_A^{}}{\partial M_\pi^{}} \equiv
\frac{\partial g_A^{}}{\partial M_*^{}}
\frac{\partial M_*^{}}{\partial M_\pi^{}},
\end{equation}
where
\begin{equation}
M_*^{} \equiv \frac{M_\pi^{}}{4\pi F_\pi^{}},
\end{equation}
and
\begin{equation}
\frac{\partial M_*^{}}{\partial M_\pi^{}} = 
\frac{1}{4\pi F_\pi^{}} \left(1 - \frac{M_\pi^{}}{F_\pi^{}} \frac{\partial F_\pi^{}}{\partial M_\pi^{}} \right),
\label{dMstar_dMpi}
\end{equation}
in terms of which
\begin{equation}
g_A^{} = 1.273(19), \qquad
\left.\frac{\partial g_A^{}}{\partial M_*^{}}\right|_{M_\pi^\mathrm{ph}} = -0.08(24),
\label{gA_extr}
\end{equation}
were obtained from an extrapolation using the complete NNLO chiral expression, with and without inclusion of the N3LO contact
terms~\cite{Bernard:2006te}. It should be noted that unlike the determination of $g_A$ itself, the value of
$\partial g_A/\partial M_*$ does depend significantly on the choice
of extrapolation of the lattice QCD data. For instance, significantly larger values can be obtained by means of linear or quadratic
extrapolations in $M_*$. However, we shall here rely on the chiral NNLO result~(\ref{gA_extr}), in particular as it was
found to show good convergence of the chiral expansion~\cite{Chang:2018uxx}. 

The dependence of $F_\pi$ on $M_\pi$ was not yet obtained in
Ref.~\cite{Chang:2018uxx}. We recall that Ref.~\cite{Berengut:2013nh} provided
\begin{equation}
\left.\frac{\partial F_\pi^{}}{\partial M_\pi^{}}\right|_{M_\pi^\mathrm{ph}} = 0.066(16),
\label{Fpi_der_K}
\end{equation}
which was used in Ref.~\cite{Epelbaum:2013wla}. This should be compared with the 
sub-leading order ChPT result
\begin{equation}
\frac{\partial F_\pi^{}}{\partial M_\pi^{}} = \frac{M_\pi^{}}{8\pi^2 F} \bar\ell_4^{},
\label{Fpi_der_ell}
\end{equation}
where $F \simeq 86.2$~MeV denotes $F_\pi$ in the chiral limit, and $\bar{\ell}_4 = 4.3(3)$ from 
the review~\cite{Bijnens:2014lea}. We note that Eq.~(\ref{Fpi_der_ell}) gives a number comparable with (though
slightly larger than) Eq.~(\ref{Fpi_der_K}). These values are also consistent with the most
recent FLAG lattice QCD determination~\cite{Aoki:2019cca}.

Using the isospin-averaged pion mass $M_\pi = 138.03$~MeV and $F_\pi = 92.1$~MeV, 
we find
\begin{equation}
\left.\frac{\partial M_*^{}}{\partial M_\pi^{}}\right|_{M_\pi^\mathrm{ph}} = 0.078(2) \: \mathrm{l.u.},
\label{dMstar_dMpi_num}
\end{equation}
from Eq.~(\ref{dMstar_dMpi}), for an inverse spatial lattice spacing of $a^{-1} \equiv 100$~MeV.
So far, from Eq.~(\ref{eq:x2}) and~(\ref{Fpi_der_K}), we have
\begin{equation}
x_2^{} = -0.050(12) \: \mathrm{l.u.} + \frac{1}{2 F_\pi^{}} 
\left.\frac{\partial g_A^{}}{\partial M_\pi^{}}\right|_{M_\pi^\mathrm{ph}},
\end{equation}
where we can now use the chiral lattice QCD extrapolations~(\ref{gA_extr})
together with Eq.~(\ref{dMstar_dMpi_num}). This gives
\begin{equation}
\left.\frac{\partial g_A^{}}{\partial M_\pi^{}}\right|_{M_\pi^\mathrm{ph}} = -0.006(19) \: \mathrm{l.u.},
\end{equation}
such that we finally obtain
\begin{equation}
x_2^{} = -0.053(16) \: \mathrm{l.u.},
\label{eq:x2val}
\end{equation}
from Eq.~(\ref{eq:x2}),
which is compatible with the range $x_2 = -0.056 \ldots 0.008 \: \mathrm{l.u.}$ used in Ref.~\cite{Epelbaum:2013wla}. 
However, in contrast to the earlier determination of $x_2$, we can now pin it down with a definite sign up to $\simeq 3.3\sigma$. 

\begin{table*}[t]
  \caption{Available lattice QCD results for the deuteron and dineutron binding energies obtained from the plateau method
    along with the resulting values of the inverse scattering lengths calculated from the LETs at NLO. The uncertainties in
    the energies are taken from the correponding papers. The first error of $a^{-1}_{s,t}$ reflects the uncertainty of the
    lattice results for the binding energies used as input (for $M_\pi = 300$, $390$ and $510$~MeV,
    the different lattice errors for the binding energies have been added in quadrature), while the second one corresponds
    to the uncertainty of the LETs estimated as explained in Ref.~\cite{Baru:2016evv}.
\label{data}}
\smallskip
\begin{tabular*}{\textwidth}{@{\extracolsep{\fill}}rrrrr}
\hline
\hline
\noalign{\smallskip}
 &   $M_\pi = 300$ MeV \cite{Yamazaki:2015asa}&     $M_\pi = 390$ MeV \cite{Beane:2011iw} &    $M_\pi = 450$ MeV \cite{Orginos:2015aya} & $M_\pi=510$ MeV \cite{Yamazaki:2012hi} 
\smallskip
 \\
\hline
\hline
\multicolumn{5}{l}{The $^3$S$_1$ channel } \\
$B_d$ [MeV] &  $14.5(0.7)({}^{+2.4}_{-0.8})$  & $11(05)(12)$ & $14.4({}^{+3.2}_{-2.6})$ &  $11.5(1.1)(0.6)$ \\  [2.5pt]
$a_t^{-1}$ [fm$^{-1}$] & $0.422({}^{+0.024}_{-0.011})({}^{+0.008}_{-0.006})$  & $0.400({}^{+0.119}_{-0.400})({}^{+0.018}_{-0.011})$  & $0.448({}^{+0.031}_{-0.029})({}^{+0.042}_{-0.093})$  & $0.419({}^{+0.015}_{-0.016})({}^{+0.077}_{-0.019})$ \\  
[4pt]
\hline
\multicolumn{5}{l}{The $^1$S$_0$ channel } \\
$B_{nn}$ [MeV] & $8.5(0.7)({}^{+1.6}_{-0.5})$  & $7.1(5.2)(7.3)$ & $12.5({}^{+3.0}_{-5.0})$ &  $7.4(1.3)(0.6)$ \\   [2.5pt]
$a_s^{-1}$ [fm$^{-1}$] & $0.335({}^{+0.023}_{-0.013})({}^{+0.009}_{-0.006})$   & $0.324({}^{+0.106}_{-0.324})({}^{+0.019}_{-0.009})$ & $0.324({}^{+0.030}_{-0.064})({}^{+0.055}_{-0.019})$  &  $0.337({}^{+0.021}_{-0.025})({}^{+0.183}_{-0.016})$ \\ 
[4pt]
\hline
\hline
\end{tabular*}
\end{table*}

Having fixed the hadronic input parameters, we now consider the leading order four-nucleon contact interactions that
can be mapped on the derivatives of the inverse singlet and triplet neutron-proton scattering lengths,
\begin{equation}
\label{eq:defAst}  
\bar A_{s,t}^{} = \frac{\partial a_{s,t}^{-1}}{\partial M_\pi^{}}\biggl|_{M_\pi^{\rm ph}}~.
\end{equation}
Earlier,  modelling based on resonance saturation~\cite{Epelbaum:2001fm} was used
to get a handle on these quantities, as discussed in detail in Ref.~\cite{Berengut:2013nh}. Here, we attempt
to fix $\bar A_{s,t}^{}$ from available lattice QCD data, which should be the method of choice. Before doing so, some
words of caution are in order. The situation with lattice QCD simulations in the nucleon-nucleon (NN) sector
is at present highly controversial. Fully dynamical  simulations at unphysically heavy pion masses, carried out
by the NPLQCD Collaboration and Yamazaki et al.~and based on the standard
approach to extract the ground state energy by fitting plateaus of the correlation functions
find more attraction in both the $^1$S$_0$ and $^3$S$_1$ channels at heavy pion masses than for physical
pion masses, see~\cite{Beane:2011iw,Yamazaki:2015asa,Orginos:2015aya,Beane:2013br,Yamazaki:2012hi}.
These results contradict the findings of the HAL QCD Collaboration using a (scheme-dependent) potential
at the intermediate stage of extracting NN observables. This group finds no bound states in both
S-wave channels for pion masses ranging from $469$ to $1171$~MeV \cite{Inoue:2011ai}. The HAL QCD Collaboration
has already carried out simulations at the physical point, but the results for the nonstrange channels have, as far
as we know, not been released yet.  The HAL QCD Colalboration has criticized the direct method by pointing
out the danger of observing fake plateaus \cite{Iritani:2017rlk,Aoki:2017byw}, see, however, the response of
the NPLQCD Collaboration in Ref.~\cite{Beane:2017edf}. The HAL QCD approach has also been criticized e.g.~in
Refs.~\cite{Birse:2012ph,Haidenbauer:2019utu}. The weakest point of
this method seems to be its reliance on the derivative expansion, whose convergence is not clear
a priori. Interestingly, the recent lattice QCD study in the strangeness $S= -2$ two-baryon sector
by the CERN-Mainz group \cite{Francis:2018qch} using a superior distillation method  finds for $M_\pi = 960$~MeV the H-dibaryon
energy perfectly consistent with HAL QCD, but in a strong disagreement with
the NPLQCD result.

\begin{figure*}[h]
\includegraphics[width=\textwidth,keepaspectratio,angle=0,clip]{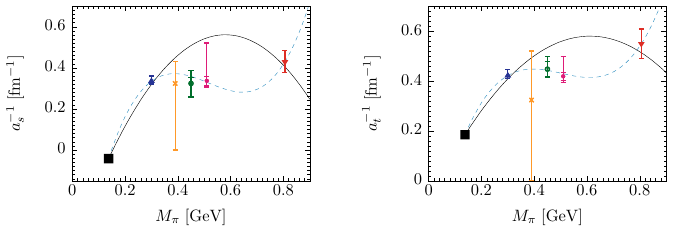}
\caption{Pion mass dependence of the inverse scattering lengths $a_{s,t}^{-1}$ as predicted
  by lattice-QCD calculations of Refs.~\cite{Yamazaki:2015asa,Beane:2011iw,Orginos:2015aya,Yamazaki:2012hi,Beane:2013br}. 
  Black squares show the experimental values at the physical point. Black solid and blue dashed lines are the quadratic and
  cubic interpolations. 
\label{fig:fit}}
\end{figure*}

Recently, in Ref.~\cite{Baru:2015ira} the use of low-energy theorems (LETs) to reconstruct the energy dependence
of the NN scattering amplitude in a large kinematical domain  from a single observable (e.g.~binding energy,
scattering length, effective range) at a given fixed value of the pion mass was proposed. The method relies on
the dominance of the one-pion exchange (OPE) at large distances, which governs the near-threshold energy
dependence of the scattering amplitude. At the physical point,  LETs are known to work accurately in
the $^3$S$_1$ channel in line with the strong tensor interaction induced by the OPE,
while less accurately in the $^1$S$_0$ partial wave, where the OPE potential is very weak \cite{Cohen:1998jr}.
Notice that this approach employs the lattice QCD results to determine the strength of the OPE potential at
unphysical values of $M_\pi$ and  does not rely on the chiral expansion. At heavy pion masses $M_\pi \sim M_\rho$,
the LETs loose their predictive power, and the approach becomes equivalent to the effective range expansion.
In \cite{Baru:2015ira},  the LETs were used to test the linear interpolation 
of $M_\pi r$ as function of $M_\pi$ between $M_\pi^{\rm ph}$ and $M_\pi \simeq 800$~MeV conjectured
by the NPLQCD Collaboration \cite{Beane:2013br}. These study was restricted to the $^3$S$_1$ channel.  
The assumed linear interpolation was indeed found to be consistent with the available lattice QCD
results for the deuteron binding energy obtained using the plateau method, see Fig.~6 of Ref.~\cite{Baru:2015ira}. 
In \cite{Baru:2016evv} the LETs were applied to test the consistency between the bound state
energies and phase shifts obtained using the L\"uscher method by  the NPLQCD
Collaboration at $M_\pi = 450$~MeV \cite{Orginos:2015aya} in both the $^1$S$_0$ and $^3$S$_1$ channels.  
It was found that the NPLQCD phase shifts are inconsistent with their own results
for the deuteron and dineutron energies. The inconsistency was later re-emphasized by the HAL QCD
Collaboration using the effective range expansion \cite{Iritani:2017rlk,Aoki:2017byw}. 

With these drawbacks and inconsistencies in mind, we nevertheless go forward and analyze the
available lattice-QCD results for the deuteron and dineutron binding energy of
Refs.~\cite{Beane:2011iw,Yamazaki:2015asa,Orginos:2015aya,Beane:2013br,Yamazaki:2012hi}, which
seem to be  mutually consistent, in order to extract the quantities $\bar A_s$ and $\bar A_t$.
The best way to extract $\bar A_{s,t}$ is to use the LETs to compute the inverse scattering
lengths from the binding energies at the corresponding values of $M_\pi$, and to perform a subsequent interpolation.  
In table~\ref{data}, we collect the binding energies of the deuteron and dineutron states
from the calculations of Refs.~\cite{Yamazaki:2015asa,Beane:2011iw,Orginos:2015aya,Yamazaki:2012hi}
and the resulting values of the inverse scattering lengths.
In addition to these results, we also include the direct NPLQCD determination of the scattering lengths
at $M_\pi \simeq 806$~MeV from Ref.~\cite{Beane:2013br}
\begin{equation}
a_t^{-1} = 0.549(59) \mbox{fm}^{-1}, \quad 
a_s^{-1} = 0.429(54) \mbox{fm}^{-1}.
\end{equation}
To perform the interpolation between these five  points and the experimental values of
the inverse scattering lengths, see Fig.~\ref{fig:fit}, we use a simple quadratic {\it ansatz}
\begin{align}
a_{s,t}^{-1} (M_\pi ) = (a_{s,t}^{\rm ph})^{-1} + a (M_\pi - M_\pi^{\rm ph})  + b (M_\pi - M_\pi^{\rm ph})^2,
\end{align}
where the coefficients $a$, $b$ are determined from a least square fit to the available values
of $a_{s,t}^{-1} $ at heavier-than-physical pion masses. The results of the fits are shown by the
solid (black) lines in Fig.~\ref{fig:fit}. With $\chi^2/N_{\rm DOF} = 4.3$ and $2.4$ in the singlet and triplet channels,
respectively, the quality of the fit is not really good. Using a third-degree polynomial leads
to the results shown by the dotted (blue) lines, but the $\chi^2/N_{\rm DOF} $ well below $1$ indicates overfitting.
The large values of $\chi^2/N_{\rm DOF} $ with a quadratic fit are likely due to underestimated
error bars of lattice-QCD results, especially the ones from \cite{Yamazaki:2015asa} at $M_\pi = 300$~MeV. 
In any case, we obtain  $\bar A_s = 0.54$ and  $\bar A_t = 0.33$ based on the quadratic
interpolation. The cubic extrapolation yields $\bar A_s = 0.78$ and  $\bar A_t = 0.49$, and we take the
difference as an estimation of the uncertainty in $\bar A_{s,t}$.   
To summarize, the various available determinations of $\bar A_{s,t}$ are:
\begin{align}
& \mbox{The original estimation in Refs.~\cite{Epelbaum:2013wla,Berengut:2013nh}:} 
\nonumber \\ 
& \qquad \bar A_s = 0.29^{+0.25}_{-0.23}, \quad \bar A_t = -0.18^{+0.10}_{-0.10},\\
\label{Kad}
& \mbox{LO chiral EFT of Ref.~\cite{Behrendt:2016nql}:} 
\nonumber \\ 
& \qquad \bar A_s = 0.50(23), \quad \bar A_t = -0.12(08), \\
\label{latt}
& \mbox{Interpolation of lattice QCD data of Refs.~\cite{Beane:2011iw,Yamazaki:2015asa,Orginos:2015aya,Beane:2013br,Yamazaki:2012hi}:} 
\nonumber \\ 
& \qquad \bar A_s = 0.54(24), \quad \bar A_t =  0.33(16). 
\end{align}
The LO chiral EFT result in Eq.~(\ref{Kad}) corresponds to an renormalizable expression for the the scattering
amplitude with the static  one-pion exchange, and the uncertainty is estimated from the cutoff variation
from $\Lambda = 600$~MeV to infinity. The positive sign of $\bar A_t$ in Eq.~(\ref{latt}) is consistent
with a stronger attraction in the deuteron channel at heavy $M_\pi$. HAL QCD results would presumably
yield a negative value of $\bar A_t$.  In what follows, we will use the values collected in Eq.~(\ref{latt}).


\section{The Hoyle state in stellar nucleosynthesis \label{sec_stellar}}

The stellar synthesis of elements heavier than $^{4}$He is complicated by the fact that no stable nucleus exists for $A = 8$, at least
for the physical values of the fundamental constants. In the absence of stable $^{8}$Be nuclei, helium fusion instead takes place
through the triple-alpha reaction 3($^{4}$He) $\to$ $^{12}$C + $\gamma$, which requires a number of intermediate steps.
The first step is $^{4}$He + $^{4}$He $\leftrightarrow$ $^{8}$Be, whereby a transient equilibrium population of $^{8}$Be is maintained
in the stellar core. It should be noted that the unstable $^{8}$Be resonance decays back into two alpha particles with a half-life
of $\sim 10^{-16}$~s. The reaction rate for the formation of $^{8}$Be is controlled by the energy difference
\begin{equation}
\Delta E_b^{} \equiv E_8^{} - 2 E_4^{}, 
\end{equation}
where $E_4^{}$ and $E_8^{}$ denote the ground states of $^{4}$He and $^{8}$Be, respectively. Though $^{8}$Be is short-lived, 
a sufficiently large transient $^{8}$Be population in stellar cores is formed to allow for the second step 
$^{8}$Be + $^{4}$He $\to$ $^{12}$C in the triple-alpha process.

In stellar cores composed primarily of helium, the non-resonant
reaction proceeds too slowly to explain the observed abundances of carbon and oxygen in the universe. However, as the $^{12}$C nucleus
possesses an excited $^{12}$C($0_2^+$) state (known as the Hoyle state) with an empirical excitation energy of $7.6444$~MeV, the
reaction can also proceed in a resonant manner, which greatly enhances the triple-alpha reaction rate. We define
\begin{equation}
\Delta E_h^{} \equiv E_{12}^\star - E_8^{} - E_4^{},
\end{equation}
which controls the reaction rate for the second step
$^{8}$Be + $^{4}$He $\leftrightarrow$ $^{12}$C($0_2^+$), where $E_{12}^\star$ is the (total) energy of the Hoyle state resonance. 
The energy scale $E_R$ which controls the resonant triple-alpha reaction is then
\begin{equation}
E_R^{} \equiv \Delta E_b^{} + \Delta E_h^{} = E_{12}^\star - 3 E_4^{},
\label{ER_def}
\end{equation}
which is empirically known (in our universe) to be $E_R^\mathrm{ph} = 379.47(18)$~keV. Note that this is much smaller than the binding energies of the 
nuclei participating in the triple-alpha reaction, which are $\simeq 28$~MeV for $^{4}$He and $\simeq 92$~MeV for $^{12}$C. 

For a stellar plasma 
at temperature $T$, the reaction rate $r_{3 \alpha}$ for fusion of three alpha particles via the 
ground state of $^8$Be and the Hoyle state of $^{12}$C is 
\begin{equation}
\label{rate}
r_{3 \alpha}^{} = 3^{\frac{3}{2}} N_\alpha^3 
\left( \frac{2 \pi \hbar^2}{|E_4^{}| k_B^{} T} \right)^3 
\frac{\Gamma_\gamma^{}}{\hbar} \, \exp \left( -\frac{E_R^{}}{k_B^{} T} \right),  
\end{equation}
where $N_\alpha$ is the number density of alpha particles, and $k_B$ is the Boltzmann constant. It should be noted how the
exponential dependence on $E_R$ and $T^{-1}$ arises, as the observation that the rate of stellar carbon production is exponentially sensitive
to $E_R$ is central to the anthropic picture of the triple-alpha process, for discussions on this issue see~\cite{kragh,Meissner:2014pma}.
For non-resonant reactions, the corresponding factor is given by the convolution of
Coulomb barrier penetration with a thermal distribution of particle velocities, which gives a $\sim T^{-1/3}$ dependence on temperature. For
resonant reactions a fixed energy is singled out, in this case given by Eq.~(\ref{ER_def}), which leads to the exponential dependence of Eq.~(\ref{rate}).
It should be noted that $E_R$ is clearly the dominant control parameter of the triple-alpha process, in comparison with the linear dependence
on the radiative width $\Gamma_\gamma \simeq 0.0037$~eV of the Hoyle state. Still, $\Gamma_\gamma$ should be sufficiently large to allow
for the radiative decay of the Hoyle state to be competitive with fragmentation into $^8$Be and $^4$He. The radiative decay proceeds either
through $^{12}$C($0_2^+$) $\to$ $^{12}$C($0_1^+$) + $\gamma$, or
$^{12}$C($0_2^+$) $\to$ $^{12}$C($2_1^+$) + $\gamma$, after which the $^{12}$C($2_1^+$) decays to the ground state
$^{12}$C($0_1^+$) by emission of a second photon. In practice, this two-step $E2$ process is more efficient than the direct $M0$ decay, which
is highly suppressed. Interestingly, the channel $^{12}$C($0_2^+$) $\to$ $^{12}$C($2_1^+$) + $\gamma$ is strongly enhanced compared to
the single-particle Weisskopf rate (often referred to as ``strongly collective behavior''), which is correctly predicted by recent lattice EFT calculations.

We define
\begin{equation}
\delta E_R^{} \equiv E_R^{} - E_R^\mathrm{ph},
\end{equation}
when the energy of the Hoyle state is shifted from its physical value. Clearly, for $\delta E_R > 0$ (the Hoyle state energy is increased), 
the rate of carbon production (at constant stellar temperature $T$) is decreased. In order to generate sufficient energy to counteract gravitation, 
the stellar core must increase $T$ in order to compensate for the reduction in $r_{3 \alpha}$. It should be noted that the production of $^{12}$C
via the triple-alpha reaction competes with the destruction of $^{12}$C by the formation of $^{16}$O through
$^{12}$C+$^{4}$He $\to$ $^{16}$O$+\gamma$, such that even a small change in $T$ could lead to a C/O abundance which is very different from that
observed. However, the $^{16}$O nucleus has a state with excitation energy $7.1187$~MeV, which is below the threshold energy of
the $^{12}$C+$^{4}$He system, which is  $7.1616$~MeV above the ground state of $^{16}$O. Hence, the formation of $^{16}$O is
a non-resonant process, which nevertheless depends sensitively on $T$ because of the sizeable Coulomb barrier.

On a phenomenological level, a variation of $E_R$ thus leads to a change in the relative importance of the competitive processes 
by which $^{12}$C is produced, and destroyed by further processing into $^{16}$O (and heavier alpha nuclei). Hence, for a sufficiently large
positive $\delta E_R$, a regime is encountered in which little $^{12}$C and $^{16}$O remains after stellar nucleosynthesis, with most material
having been processed into $^{24}$Mg and $^{28}$Si. Conversely, for negative $\delta E_R$, stellar core temperatures during helium burning
are substantially lower, leading to end products with plentiful $^{12}$C but relatively little $^{16}$O. However, the latter point turns out to be
sensitive to the initial stellar metallicity. Also, when the Hoyle state energy is lowered ($\delta E_R < 0$), one should also consider
the sensitivity
\begin{equation}
\Xi_T^{} \equiv \frac{T}{r_{3\alpha}^{}} \frac{d r_{3\alpha}^{}}{dT} =
-3 + \frac{E_R^{}}{k_B^{} T},
\end{equation}
where the stability of the star requires that the triple-alpha reaction rate (and hence the energy production) increase as $T$ increase, such
that $\Xi_T > 0$. Hence, $E_R$ should satisfy
\begin{equation}
E_R^{} > 3k_B^{} T,
\label{ER_constraint}
\end{equation}
which places a lower bound on the permissible values of the Hoyle state energy. However, as stellar cores have roughly $k_B T \simeq 10$~keV
during helium burning, this bound is an order to magnitude smaller than the observed value of $E_R \simeq 380$~keV.

Recently, comprehensive simulations of stellar nucleosynthesis in massive stars that eventually explode as supernovae have
become available~\cite{Huang:2018kok}. These studies follow stars ranging from 15~to 40~solar masses up to the stage where a degenerate iron
core is formed. The yields of various isotopes are then weighted according to the stellar mass distribution function, taken to be
$dN/dM_\odot \propto M_\odot^{-2.3}$ (for the range of stellar masses $M_\odot$ considered), which accounts for the relative scarcity of heavier stars.
For the stellar metallicity, Ref.~\cite{Huang:2018kok} considered two cases. Firstly, the low-metallicity simulations used $\mathcal{Z} = 10^{-4}$, 
which is representative of the currently known stars with the lowest observed metallicity. Secondly, the case of $\mathcal{Z} = \mathcal{Z}_\mathrm{sun}$ was studied,
where $\mathcal{Z}_\mathrm{sun} = 0.02$ denotes the observed solar metallicity. The main effect of $\mathcal{Z}$ is to alter the relative
importance of the $p$-$p$ chain and the CNO cycle, such that stars can enter the helium burning phase with different configurations.

We are now in a position to summarize the findings of Ref.~\cite{Huang:2018kok} for the range in $\delta E_R$, for which the final abundances of 
$^{12}$C and $^{16}$O exceed their initial values. These can be expressed as
\begin{equation}
\delta E_R^{(-)} \leq \delta E_R^{} \leq \delta E_R^{(+)},
\end{equation}
where the boundaries for each nucleus depend on the chosen initial stellar metallicity.
For $\mathcal{Z} = 10^{-4}$ (low metallicity), the ranges compatible with carbon-oxygen based 
life are
\begin{equation}
\mathrm{^{12}C} (\mathcal{Z} = 10^{-4}): \quad -300~\mathrm{keV} \leq \delta E_R^{} \leq 500~\mathrm{keV},
\end{equation}
and
\begin{equation}
\mathrm{^{16}O} (\mathcal{Z} = 10^{-4}): \quad -300~\mathrm{keV} \leq \delta E_R^{} \leq 300~\mathrm{keV},
\end{equation}
where for negative $\delta E_R$, sufficient $^{12}$C and $^{16}$O were produced for all values of the Hoyle state energy
compatible with the constraint~(\ref{ER_constraint}). For $\mathcal{Z} = 0.02$ (solar metallicity), the corresponding ranges were found
to be significantly narrower. Specifically, 
\begin{equation}
\mathrm{^{12}C} (\mathcal{Z} = 0.02): \quad -300~\mathrm{keV} \leq \delta E_R^{} \leq 160~\mathrm{keV},
\end{equation}
and
\begin{equation}
\mathrm{^{16}O} (\mathcal{Z} = 0.02): \quad -150~\mathrm{keV} \leq \delta E_R^{} \leq 200~\mathrm{keV},
\end{equation}
where significant carbon production could still be maintained for all negative $\delta E_R$. In stars of solar metallicity, the
production of $^{16}$O appears to be the limiting factor, as it is only possible in a roughly symmetric (though rather broad) 
envelope centered on the physical value of $E_R$. 

Previously, 
the stellar simulations of Refs.~\cite{Oberhummer:2000zj,Oberhummer:1999ab} indicated 
that sufficient abundances of both carbon and oxygen are only possible for
$\delta E_R \simeq \pm 100$~keV around the empirical value $E_R= 379.47(18)$~keV.
From the present results, we conclude that the energy of the Hoyle state is likely to be less fine-tuned than
previously thought. However, if the Hoyle state is raised by more than
$\simeq 300$~keV, the generation of sufficient oxygen would encounter difficulties. Were the Hoyle state located more than
$\simeq 500$~keV above its physical energy, the universe would also be unlikely to contain a sufficient amount of carbon.


\section{Sensitivity to small changes in the fundamental parameters \label{sec_sensitivity}}

We shall now update the ranges of variation of the light quark mass
$\delta m_q$ and the fine-structure constant $\delta \alpha_{\rm em}$, which are 
compatible with the formation of sufficient amounts of carbon and oxygen
in our universe, and thus with the existence of carbon-oxygen based life. As in Ref.~\cite{Epelbaum:2013wla}, 
we express the shift in the Hoyle state $\delta E_R$ as
\begin{align} 
\delta E_R^{} & \approx
\left. \frac{\partial E_R^{}}{\partial M_\pi^{}} \right |_{M_\pi^\mathrm{ph}} \delta M_\pi^{} + 
\left. \frac{\partial E_R^{}}{\partial \alpha_{\rm em}^{}} \right |_{\alpha_{\rm em}^\mathrm{ph}} 
\delta \alpha_{\rm em}^{} \\
& \equiv Q_{\rm q}^{}(E_R^{}) \left(\frac{\delta m_q^{}}{m_q^{}}\right) 
+ Q_{\rm em}^{}(E_R^{}) 
\left(\frac{\delta \alpha_{\rm em}^{}}{\alpha_{\rm em}^{}}\right),
\nonumber
\end{align}
for $| \delta m_q/m_q | \ll 1$ and
$| \delta \alpha_{\rm em}/\alpha_{\rm em} | \ll 1$. We shall first consider the effects
of varying $m_q$, for which we have
\begin{align}
Q_{\rm q}^{}(E_R^{}) \equiv 
\left. \frac{\partial E_R}{\partial M_\pi^{}} \right |_{M_\pi^\mathrm{ph}}
K_{M_\pi}^q M_\pi^{},
\end{align}
and we recall that $K_{M_\pi}^q =
0.494^{+0.009}_{-0.013}$~\cite{Berengut:2013nh}. As in the NLEFT calculation of 
Ref.~\cite{Epelbaum:2013wla}, we find
\begin{align}
\left. \frac{\partial E_R}{\partial M_\pi^{}} \right |_{M_\pi^\mathrm{ph}}
= -0.572(19) \, \bar A_s^{} - 0.933(15) \, \bar A_t^{} + 0.068(7),
\label{ER_shift}
\end{align}
where the only change from Ref.~\cite{Epelbaum:2013wla} is in the constant term, which has been recalculated
using the updated values of $x_1$ and $x_2$. The numbers in parentheses denote Monte Carlo uncertainties, and as in
Ref.~\cite{Epelbaum:2013wla}, the relatively small additional errors due to the uncertainties of $x_1$ and $x_2$ have been
neglected. As such uncertainties are much reduced here, this simplification is better justified.

\begin{figure*}[t]
\centering
\includegraphics[width = .9\textwidth]{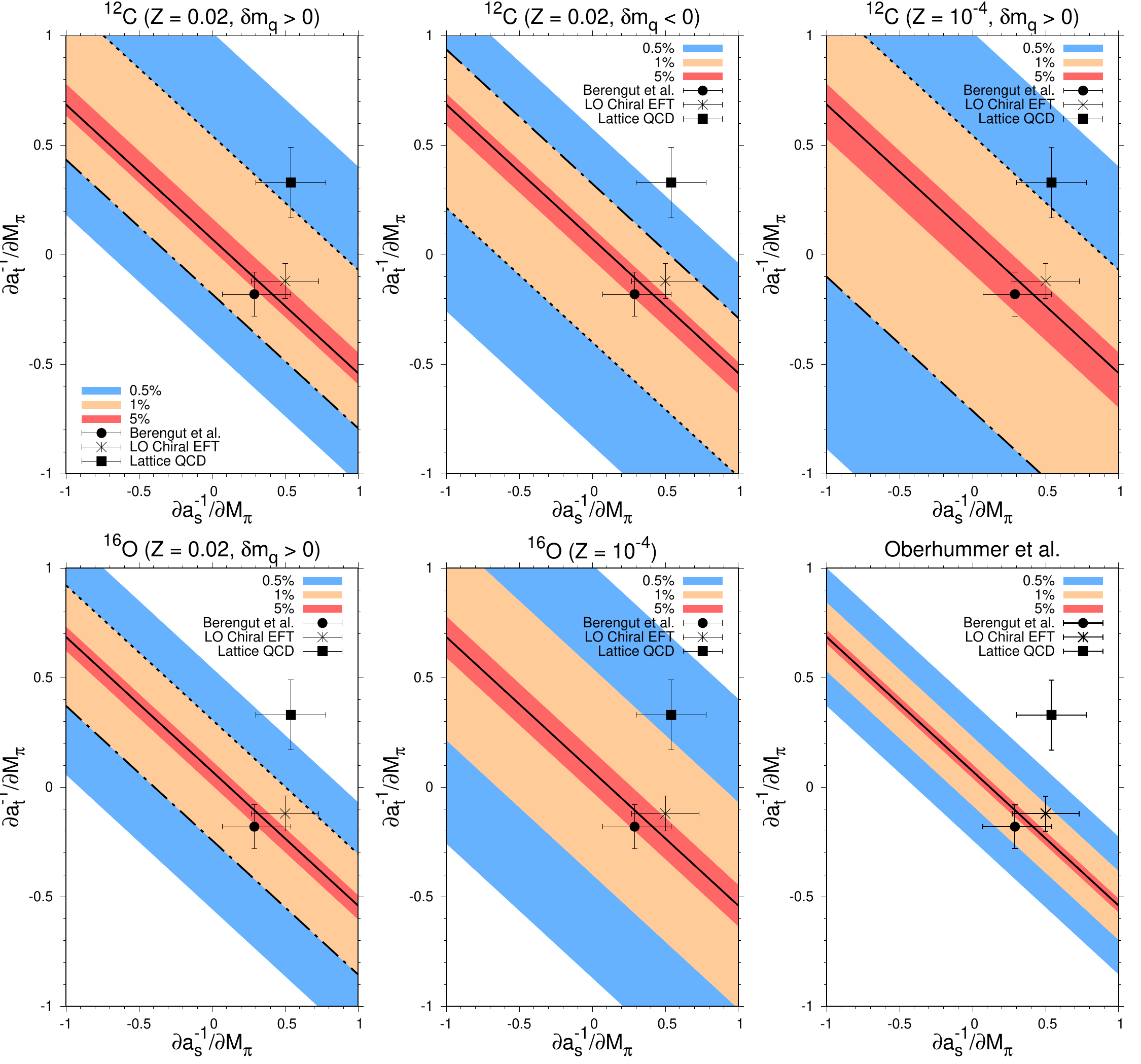}
\caption{``Survivability plots'' based on the stellar simulations of~\cite{Oberhummer:2000zj,Oberhummer:1999ab} (Oberhummer {\it et al.}, bottom right plot) 
and~\cite{Huang:2018kok,Adams:2019kby} (all other plots). The data points with horizontal and vertical error bars indicate estimates of 
where our universe is located. On the solid black lines, $E_R$ (and hence the triple-alpha reaction) is independent 
of variations in $m_q$. The dot-dashed black lines correspond to the equalities~(\ref{ER_plus_dmq_plus}) and~(\ref{ER_plus_dmq_minus}), which
mark the upper bound $\delta E_R^{(+)}$ for $|\delta m_q | / m_q = 1\%$. Similarly, the dotted black lines correspond to the
equalities~(\ref{ER_minus_dmq_plus}) and~(\ref{ER_minus_dmq_minus}), which mark the lower bound $\delta E_R^{(-)}$. For 
the case of $^{12}$C and $\mathcal{Z} = 0.02$, both $\delta m_q > 0$ and $\delta m_q < 0$ are given, for the other cases only 
$\delta m_q > 0$ is shown. 
\label{fig_survivability}}
\end{figure*}

The boundaries of the envelope where a sufficient abundance of carbon or oxygen is maintained are then
\begin{align}
\delta E_R^{(-)} \leq Q_{\rm q}^{}(E_R^{})
\left(\frac{\delta m_q^{}}{m_q^{}} \right)
\leq \delta E_R^{(+)},
\label{envelope_mq}
\end{align}
where the earlier stellar nucleosynthesis calculations of Refs.~\cite{Oberhummer:2000zj,Oberhummer:1999ab} indicated an overall
bound of $|\delta E_R^{(-)}|$ $=$ $|\delta E_R^{(+)}|$ $=$ $100$~keV from {\it ad hoc} variations of the Hoyle state energy.
The present situation is slightly more involved, as the upper and lower boundaries are not symmetric,
and moreover the boundaries for $^{12}$C and $^{16}$O are different, an additional factor being the 
(initial) metallicity of the star under consideration. On the one hand,
for $E_R$ to not increase beyond the permissible range, we require that
\begin{align}
& -0.572(19) \, \bar A_s^{} - 0.933(15) \, \bar A_t^{} + 0.068(7) 
\nonumber \\
& \leq \frac{\delta E_R^{(+)}}{K_{M_\pi}^q M_\pi^{}}
\left(\frac{|\delta m_q^{}|}{m_q^{}} \right)^{-1},
\quad
\delta m_q^{} > 0,
\label{ER_plus_dmq_plus}
\end{align}
for positive shifts in $m_q$ such that $\delta m_q \to |\delta m_q |$, and
\begin{align}
& -0.572(19) \, \bar A_s^{} - 0.933(15) \, \bar A_t^{} + 0.068(7) 
\nonumber \\
& \geq -\frac{\delta E_R^{(+)}}{K_{M_\pi}^q M_\pi^{}}
\left(\frac{|\delta m_q^{}|}{m_q^{}} \right)^{-1},
\quad
\delta m_q^{} < 0,
\label{ER_plus_dmq_minus}
\end{align}
for negative shifts in $m_q$ such that $\delta m_q \to -|\delta m_q |$. On the other hand, 
for $E_R$ to not decrease too much, we should have
\begin{align}
& -0.572(19) \, \bar A_s^{} - 0.933(15) \, \bar A_t^{} + 0.068(7) 
\nonumber \\
& \geq \frac{\delta E_R^{(-)}}{K_{M_\pi}^q M_\pi^{}}
\left(\frac{|\delta m_q^{}|}{m_q^{}} \right)^{-1},
\quad
\delta m_q^{} > 0,
\label{ER_minus_dmq_plus}
\end{align}
for positive shifts in $m_q$, and
\begin{align}
& -0.572(19) \, \bar A_s^{} - 0.933(15) \, \bar A_t^{} + 0.068(7) 
\nonumber \\
& \leq -\frac{\delta E_R^{(-)}}{K_{M_\pi}^q M_\pi^{}}
\left(\frac{|\delta m_q^{}|}{m_q^{}} \right)^{-1},
\quad
\delta m_q^{} < 0,
\label{ER_minus_dmq_minus}
\end{align}
for negative shifts in $m_q$. Hence, the values of $\bar A_s$ and $\bar A_t$ compatible with
a given variation in $m_q$ are given by the regions enclosed by Eqs.~(\ref{ER_plus_dmq_plus}) 
and~(\ref{ER_minus_dmq_plus}) for $\delta m_q^{} > 0$, and by Eqs.~(\ref{ER_plus_dmq_minus}) 
and~(\ref{ER_minus_dmq_minus}) for $\delta m_q^{} < 0$.

The constraints on $\bar A_s$ and $\bar A_t$ due to the conditions~(\ref{ER_plus_dmq_plus}) through~(\ref{ER_minus_dmq_minus})
are illustrated by the shaded bands in Fig.~\ref{fig_survivability}. These bands cover the values of $\bar A_s$ and $\bar A_t$ consistent
with the ability of stars to produce $^{12}$C and $^{16}$O, when $m_q$ is varied by $0.5$\%, $1$\% and $5$\%. As an example, 
given the boundaries for $^{12}$C production in stars with solar metallicity ($\mathcal{Z} = 0.02$), the interpolated lattice QCD result
is compatible with a $\simeq 0.8\%$ increase in $m_q$, beyond which $E_R$ is decreased too much. Conversely, beyond
a $\simeq 0.4\%$ decrease in $m_q$, $E_R$ is increased too much. For stars with solar metallicity, 
the production of $^{16}$O is clearly the most heavily constraining factor, although low-metallicity stars ($\mathcal{Z} = 10^{-4}$) 
allow for much greater variation of $m_q$. This value of the metallicity corresponds to the metal-poorest stars observed in the universe.
In such metal-poor stars, the chiral EFT determinations of $\bar A_s$
and $\bar A_t$ suggest that changes of $\sim 5\%$ in $m_q$ are
permissible. However, from the interpolated lattice QCD results, the allowed range in $m_q$ is strongly reduced to a mere $\simeq 0.8\%$, largely
because of the positive value of $\bar A_t$. 

Finally, we note that the effect of shifts in the electromagnetic fine-structure constant lead to the
constraint 
\begin{align}
\frac{|\delta \alpha_{\rm em}^{}|}{\alpha_{\rm em}^{}}
\leq \frac{|\delta E_R^{}|}{Q_{\rm em}^{}(E_R^{})},
\label{envelope_alpha}
\end{align}
where $Q_{\rm em}(E_R) = 3.99(9)$~MeV was determined in the NLEFT calculation of Ref.~\cite{Epelbaum:2013wla}. With the
bound $|\delta E_R| = 100$~keV~\cite{Oberhummer:2000zj,Oberhummer:1999ab}, this is compatible with a $\simeq 2.5\%$ shift
in $\alpha_{\rm em}$. With the much more relaxed bound $|\delta E_R| = 300$~keV due to the production of $^{16}$O in stars 
with $\mathcal{Z} = 10^{-4}$, this tolerance is significantly increased to $\simeq 7.5\%$.


\section{Discussion \label{sec_discussion}}

We have reconsidered the sensitivity of the triple-alpha process with respect to shifts in the fundamental parameters of nature,
especially the light quark mass $m_q$ and the electromagnetic fine-structure constant $\alpha_{\rm em}$. Our knowledge of
the quark-mass dependence of the hadronic parameters in the nuclear Hamiltonian has improved significantly, in particular
with respect to the nucleon axial-vector coupling $g_A$. There, new lattice QCD data allow for an accurate chiral extrapolation which
shows good convergence. Much more detailed predictions of the effects of {\it ad hoc} variation
of the position of the Hoyle state resonance on the stellar yields of $^{12}$C and $^{16}$O have also become available. These show
that the production of $^{16}$O in low-metallicity stars is likely to be the limiting factor, although the bounds on carbon-oxygen based life
have in general become much less stringent. At the same time, much more lattice QCD data on the singlet and triplet S-wave nucleon
scattering lengths at unphysical quark masses have been produced. The current lattice QCD data appear to exclude the no-fine-tuning scenario, 
to the extent that a relatively small $\simeq 0.5\%$ shift in $m_q$ would eliminate carbon-oxygen based life from the universe. On the other hand,
such life could possibly persist up to $\simeq 7.5\%$ shifts in
$\alpha_{\rm em}$. Clearly, more reliable lattice QCD data at close-to-physical
pion masses are required to overcome the remaining uncertainties discussed in detail in Sec.~\ref{sec_Ham}. 

While we have here mostly focused on updating the nuclear, hadronic and astrophysical inputs to the EFT calculation of the fine-tuning of the
triple-alpha process, it is also of interest to perform improved $^{12}$C simulations in NLEFT.
Apart from effects of smearing of the LO operators, the EFT calculation of the triple-alpha process is essentially a LO calculation. Extending this
to higher orders would require much more detailed information on the nuclear force at unphysical pion masses, for higher orders in the EFT expansion.
However, as modern NLEFT potentials, see e.g.~\cite{Li:2018ymw},
use a combination of local and non-local smearing at LO, the description of $^{12}$C including the Hoyle state
is nevertheless expected to be improved.


\section*{Acknowledgments}
We are grateful to Evan Berkowitz for supplying the analysis of the pion mass
dependence of $g_A$.
This work is supported in part by  the DFG (Grant No. TRR110)
and the NSFC (Grant No. 11621131001) through the funds provided
to the Sino-German CRC 110 ``Symmetries and the Emergence of
Structure in QCD", by the BMBF (Grant No.  05P2015), 
by the Chinese  Academy of Sciences (CAS) President's International Fellowship Initiative (PIFI)
(grant no. 2018DM0034)  and by VolkswagenStiftung (grant no. 93562).
Computational resources for this project were provided 
by the J\"{u}lich Supercomputing Centre (JSC) at the Forschungszentrum 
J\"{u}lich  and by RWTH Aachen.



\begin{thebibliography}{99}
  
\bibitem{Hoyle:1954zz}
  F.~Hoyle,
  Astrophys.\ J.\ Suppl.\  {\bf 1} (1954) 121.
  
\bibitem{Adams:2019kby}
  F.~C.~Adams,
  Phys.\ Rept.\  {\bf 807} (2019) 1.

\bibitem{Huang:2018kok}
  L.~Huang, F.~C.~Adams and E.~Grohs,
  Astropart.\ Phys.\  {\bf 105} (2019) 13.

\bibitem{Lahde:2019npb}
  T.~A.~L\"ahde and U.-G.~Mei{\ss}ner,
  Lect.\ Notes Phys.\  {\bf 957} (2019) 1.

  
\bibitem{Epelbaum:2012iu}
  E.~Epelbaum, H.~Krebs, T.~A.~L\"ahde, D.~Lee and U.-G.~Mei{\ss}ner,
  Phys.\ Rev.\ Lett.\  {\bf 110} (2013) no.11,  112502.

\bibitem{Epelbaum:2013wla}
  E.~Epelbaum, H.~Krebs, T.~A.~L\"ahde, D.~Lee and U.-G.~Mei{\ss}ner,
  Eur.\ Phys.\ J.\ A {\bf 49} (2013) 82.


\bibitem{Hoferichter:2015dsa}
  M.~Hoferichter, J.~Ruiz de Elvira, B.~Kubis and U.-G.~Mei{\ss}ner,
  Phys.\ Rev.\ Lett.\  {\bf 115} (2015) 092301.
  
\bibitem{RuizdeElvira:2017stg}
  J.~Ruiz de Elvira, M.~Hoferichter, B.~Kubis and U.-G.~Mei{\ss}ner,
  J.\ Phys.\ G {\bf 45} (2018) no.2,  024001.

\bibitem{Hoferichter:2016ocj}
  M.~Hoferichter, J.~Ruiz de Elvira, B.~Kubis and U.-G.~Mei{\ss}ner,
  Phys.\ Lett.\ B {\bf 760} (2016) 74.
  
\bibitem{Chang:2018uxx}
  C.~C.~Chang {\it et al.},
  Nature {\bf 558} (2018)   91.

\bibitem{Bernard:2006te}
  V.~Bernard and U.-G.~Mei{\ss}ner,
  Phys.\ Lett.\ B {\bf 639} (2006) 278.

\bibitem{Berengut:2013nh}
  J.~C.~Berengut, E.~Epelbaum, V.~V.~Flambaum, C.~Hanhart, U.-G.~Mei{\ss}ner, J.~Nebreda and J.~R.~Pelaez,
  Phys.\ Rev.\ D {\bf 87} (2013)   085018.
  
\bibitem{Bijnens:2014lea}
  J.~Bijnens and G.~Ecker,
  Ann.\ Rev.\ Nucl.\ Part.\ Sci.\  {\bf 64} (2014) 149.
  
\bibitem{Aoki:2019cca}
  S.~Aoki {\it et al.} [Flavour Lattice Averaging Group],
  arXiv:1902.08191 [hep-lat].

\bibitem{Epelbaum:2001fm}
  E.~Epelbaum, U.-G.~Mei{\ss}ner, W.~Gloeckle and C.~Elster,
  Phys.\ Rev.\ C {\bf 65} (2002) 044001.

\bibitem{Beane:2011iw}
  S.~R.~Beane {\it et al.}  [NPLQCD Collaboration],
  Phys.\ Rev.\ D {\bf 85} (2012) 054511.


\bibitem{Yamazaki:2015asa} 
  T.~Yamazaki, K.~i.~Ishikawa, Y.~Kuramashi and A.~Ukawa,
  Phys.\ Rev.\ D {\bf 92} (2015) 014501.
  
\bibitem{Orginos:2015aya} 
  K.~Orginos  {\it et al.}, 
  arXiv:1508.07583 [hep-lat].


\bibitem{Beane:2013br}
  S.~R.~Beane {\it et al.}  [NPLQCD Collaboration],
  Phys.\ Rev.\ C {\bf 88} (2103) 024003.


\bibitem{Yamazaki:2012hi}
  T.~Yamazaki, K.~i.~Ishikawa, Y.~Kuramashi and A.~Ukawa,
  Phys.\ Rev.\ D {\bf 86} (2012) 074514.


  \bibitem{Inoue:2011ai}
  T.~Inoue {\it et al.}  [HAL QCD Collaboration],
  Nucl.\ Phys.\ A {\bf 881} (2012) 28.

\bibitem{Iritani:2017rlk} 
  T.~Iritani {\it et al.},
  Phys.\ Rev.\ D {\bf 96} (2107) 034521.

\bibitem{Aoki:2017byw} 
  S.~Aoki, T.~Doi and T.~Iritani,
  EPJ Web Conf.\  {\bf 175} (2018) 05006.


\bibitem{Beane:2017edf} 
  S.~R.~Beane {\it et al.},
  arXiv:1705.09239 [hep-lat].

\bibitem{Birse:2012ph} 
  M.~C.~Birse,
  Eur.\ Phys.\ J.\ A {\bf 53} (2017) 223.

\bibitem{Haidenbauer:2019utu}
  J.~Haidenbauer and U.-G.~Mei{\ss}ner,
  Eur.\ Phys.\ J.\ A {\bf 55} (2019)   70.

\bibitem{Francis:2018qch} 
  A.~Francis, J.~R.~Green, P.~M.~Junnarkar, C.~Miao, T.~D.~Rae and H.~Wittig,
  Phys.\ Rev.\ D {\bf 99} (2019) 074505.

\bibitem{Baru:2015ira} 
  V.~Baru, E.~Epelbaum, A.~A.~Filin and J.~Gegelia,
  Phys.\ Rev.\ C {\bf 92} (2015) 014001.
  
\bibitem{Cohen:1998jr} 
  T.~D.~Cohen and J.~M.~Hansen,
  Phys.\ Rev.\ C {\bf 59} (1999)  13.

\bibitem{Baru:2016evv} 
  V.~Baru, E.~Epelbaum and A.~A.~Filin,
  Phys.\ Rev.\ C {\bf 94} (2016) 014001.


\bibitem{Behrendt:2016nql} 
  J.~Behrendt, E.~Epelbaum, J.~Gegelia, U.-G.~Mei{\ss}ner and A.~Nogga,
  Eur.\ Phys.\ J.\ A {\bf 52} (2016) 296.
 
\bibitem{kragh}
H.~Kragh. Arch.\ Hist.\ Ex.\ Sci. {\bf 64} (2010) 721. 

\bibitem{Meissner:2014pma}
  U.-G.~Mei{\ss}ner,
  Sci.\ Bull.\  {\bf 60} (2015)   43.
 
 \bibitem{Oberhummer:2000zj}
  H.~Oberhummer, A.~Cs\'ot\'o, and H.~Schlattl,
  Science {\bf 289}, 88 (2000).

\bibitem{Oberhummer:1999ab} 
  H.~Oberhummer, A.~Cs\'ot\'o, and H.~Schlattl,
  arXiv:astro-ph/9908247.

\bibitem{Li:2018ymw}
  N.~Li, S.~Elhatisari, E.~Epelbaum, D.~Lee, B.~N.~Lu and U.-G.~Mei{\ss}ner,
  Phys.\ Rev.\ C {\bf 98} (2018)  044002.

\end{thebibliography}
\end{document}